\documentstyle[aps,floats,twocolumn]{revtex}

\begin{document}

\draft

\twocolumn[\hsize\textwidth\columnwidth\hsize\csname@twocolumnfalse\endcsname 

\renewcommand{\thefootnote}{\fnsymbol{footnote}}

\title{First Order Transition in the Ginzburg-Landau Model}

\author{Philippe Curty and Hans Beck} 

\address{Universit\'e de Neuch\^atel, 2000 Neuch\^atel, Switzerland}

\maketitle

\begin{abstract}
  { The $d$-dimensional complex Ginzburg-Landau (GL) model is solved according to a variational method by separating phase and amplitude. The GL transition becomes first order for high superfluid density because of effects of phase fluctuations. We discuss its origin with various arguments showing that, in particular for $d = 3$, the validity of our approach lies precisely in the first order domain.}\\
{pacs: 05.70.Fh, 72.40.+k, 64.60.-i}\\
\end{abstract}
]

Ginzburg-Landau free energy functionals involving a $n$-component space and time dependent field have been widely used in order to describe different types of phase transitions in a semi-phenomenological way. The case $n=2$, corresponding to a complex field $\psi$, applies in particular to superconductivity, superfluidity, metal-insulator transitions or to magnetic systems with moments that are confined to a plane (XY moments). For such particular applications, the coefficients determining the functional can be derived from appropriate microscopic models. For superconductors, this was originally done by Gorkov \cite{gorkov} and has been refined since by numerous authors \cite{randeria}. In its simplest form, the GL functional involves a time independent field and thus describes (classical) thermodynamic and static phenomena of superconductors. For the description of dynamic phenomena, charging effects \cite{simanek} or pairing fluctuations in strong coupling superconductors \cite{randeria}, one needs the generalisation to a time dependent field.

 An important aspect of the case $n=2$ is the interplay between variations of amplitude $|\psi|$ and phase $\phi$ of the corresponding complex field $\psi = |\psi| e^{i \phi}$. In various approximate treatments, such as the mean field approximation or the Hartree decoupling of terms involving higher powers of $\psi$, the field is treated as a whole, without separating amplitude and phase. More accurate studies of the static GL problem, like the renormalisation group approach \cite{pelcovits,toner} focusing in particular on the region near the phase transition, show that the amplitude has no critical behaviour and is irrelevant at the transition. The phase transition scenario should then correspond to the one of the XY model with the same dimensionality. In the framework of the $\varepsilon$-expansion \cite{wilson,wilsonfisher}, for $d=3$, the transition is second order and seems to have the same critical exponents as the XY model. On the other hand,  Bormann and Beck \cite{bormann} have shown that amplitude fluctuations, even though not being critical by themselves, might alter the cooperative phenomenon occurring with phases, in particular in dimension 2. Like the XY model, corresponding to a fixed value of the amplitude, the 2$d$GL model can be mapped onto a Coulomb gas describing vortex-antivortex pairs. As soon as one allows for amplitude variations, these topological excitations become energetically more favorable. Taking into account gaussian amplitude fluctuations, Bormann and Beck \cite{bormann} have shown that the system may be driven into a regime where - according to Minnhagen's phase diagram \cite{minnhagen} - a first order transition replaces the usual Kosterlitz-Thouless scenario.
 
As far as superconductors are concerned, BCS theory \cite{BCS} predicts that the transition between the normal state and the superconducting state is a second order phase transition. However, it is well known that fluctuations can change the order of the transition. For example, fluctuations of the magnetic field change the GL-BCS transition to a first order transition for type I superconductors \cite{halperin}. The three state Potts model in two dimensions is an opposite example: mean field theory predicts a first order transition, whereas the actual transition is continuous. So the question of the order of the transition in the GL model, as well as the more detailed mutual influence between phase and amplitude, is still an open problem.

{The aim of this letter} is to show the reciprocal influence between phase and amplitude by separating self-consistently, from the outset, the GL functional into two parts: the amplitude part and the phase part.

According to Ginzburg-Landau theory, we define the effective hamiltonian functional
\begin{equation}
H[\psi]=\int{d^d \! r  \left[  a   t  \left|\psi \right|^2+ {b\over 2} \left|\psi \right|^4   + { \gamma \over 2} \left|  \nabla\psi \right|^2 \right] } \label{eq: hamiltonien}
\end{equation} 
where $a$, $b$ and $\gamma$ are coefficients independent of the temperature derived from a microscopic model. $t=T/T_0-1$ is the reduced temperature and $T_0$ is the mean field critical temperature. We now introduce the amplitude $|\psi|$ and the phase $\phi$ of the field $\psi = |\psi| e^{i \phi}$. On the lattice, with lattice spacing $\varepsilon$, we normalize the hamiltonian by setting $ R^2 = {|\psi|^2/ (a/b)}$, ${ \vec u} = { \vec r / \xi_0 }$, where $ \xi^2_0= \gamma /a $ is the mean field correlation length at zero temperature. The normalized hamiltonian is then:
\begin{equation}
  H[R,\phi] = k_B V_0 \left( H_R +  \sum_{i=1}^N R_i^2  \  f_{i}  \right)
\end{equation}
where $f_{i} := \sum_{j=1}^d   \left[1- \cos (\phi_i - \phi_j)  \right]$ and
   $$
     H_R  :=  \sum_{i=1}^{N}   \left[ \sigma  \left( t  R_i ^2  +   R_i^4 /2 \right)  +   \left(\nabla R_i  \right)^2 /2 \right].
   $$
where $\nabla R_i  := \sum_{j}{\vec{e}_j (R_i-R_j)}$. $j$ points to the nearest neighbours of $i$ and $\vec{e}_j$ is a unit vector in direction $j$. We have set $R_i R_j =R_i ((R_j -R_i)+R_i) \approx R_i^2$ in the XY part of the hamiltonian. Indeed, the term $R_i (R_j-R_i)$ is not important for the present discussion because our approach will be reliable when amplitude fluctuations are small compared to phase fluctuations. Only two parameters, that are in competition, remain:
$$
\sigma := \varepsilon^2 / \xi_0^2    \hspace{2 cm}           V_0 := {1 \over k_B} {a \over b} \gamma  \varepsilon^{d-2} 
$$
$\sigma$ controls amplitude fluctuations. $V_0$ corresponds to the zero temperature phase stiffness. $V_0$ is proportional to the superfluid density $a/b$ and controls the general critical behaviour. When $V_0/T_0$ is large, the critical region is small and the material behaves according to mean field theory (as BCS superconductors). When $V_0/T_0$ is of the order of 1, phase fluctuations become very large and give an upper bound for the critical temperature \cite{emery}. One has to bear in mind that $V_0$ is not a constant that only normalizes the temperature $T$ in the Boltzmann factor because of the occurrence of the temperature $t=T/T_0-1$ inside the hamiltonian $H[\psi]$.\\
Let us write the partition function in polar coordinates:
$$
Z =    \int_0^{2 \pi}  D\phi  \int_0^{\infty} DR  \ e^{ -\beta H_{eff} } 
$$
where the effective hamiltonian $H_{eff}$ is
\begin{equation}
H_{eff} =  k_B V_0 \left[  H_R + \sum_{i}    \left(  R_i ^2  f_i - {T \over V_0} \log R_i \right) \right] \label{Heff}
\end{equation}
 We keep the factor $R$ of the Jacobian, $R \ D\phi \ DR$, and take it in the exponential giving a contribution $log R$ to the potential. This incorporates the fact that small values of $R$ have a small statistical weight, due to the volume element in phase space, into the Boltzmann factor of the canonical ensemble. 
   We compute now the partition function by integrating only the phase. For this purpose, we will drop the integration on $R_i$ and search for the minimum of the free energy with respect to $R$. The partition function becomes then:
$$
 Z \approx \int_0^{2 \pi} D\phi \ e^{- \beta H_{eff}}
$$
The minimum of the free energy $F= -k_B T \log Z$ is given through the equation $\delta F/ \delta R_i=0$. Assuming that the gradient of the amplitude is zero, we have
\begin{equation}
  \sigma  \left( t  R ^2  +   R^4 \right) -  {T \over 2 V_0}  +    R^2  f({K}) =0             \label{eq:R2}
\end{equation}
where we have multiplied the equation by $R/2$. $f({K}) = \langle  f_i \rangle $ is the expectation value within the XY model of $f_i $ with a dimensionless coupling constant $K = {V_0 \over T}R^2$. Although $K$ has the same value at each lattice site and does not explicitely adapt to the vortex structure of the phase field, the average energy of the latter still determines the value of $K$ through the minimalisation of the free energy $F$. In this work, we take Monte-Carlo simulations to evaluate the function $f(K)$ which is just the energy of the XY model \cite{kleinert}. The critical temperature $T_c$ is reached when the coupling $K$ equals the critical constant $1/\omega$:
\begin{equation}
 1 = \omega K_c =  \omega \ {V_0 \over T_c} \langle R^2 \rangle (T_c)           \label{equationTc}
\end{equation}
 where $\omega = 0, \ 0.9, \ 2.2$ (for $d=1, 2, 3$ respectively) is the pure XY critical temperature on a square lattice with unit coupling constant $K=1/T$.


\begin{figure}

\let\picnaturalsize=N
\def\picsize{8.5 cm}
\def\picfilename{figGL1.eps}
\ifx\nopictures Y\else{\ifx\epsfloaded Y\else\input epsf \fi
\let\epsfloaded=Y
\centerline{\ifx\picnaturalsize N\epsfxsize \picsize\fi \epsfbox{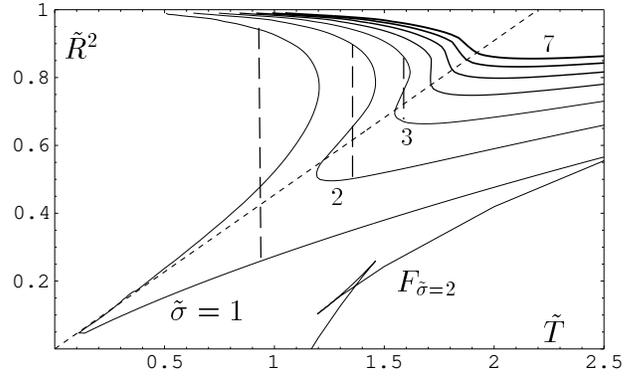}}}\fi

\caption{{\bf  Reduced mean amplitude as a function of the reduced temperature for $\bf d = 3$ for different values of $\tilde \sigma$.} { The slanting dashed line shows the XY transition ending at $\tilde T_c = \omega_{3d} = 2.2 $ ($\tilde \sigma \rightarrow \infty$). The vertical dashed lines indicate the temperature where the first order transition arises. The free energy $F_{\tilde{\sigma}=2}$ for $\tilde{\sigma} = 2 $ is also shown (in arbitrary units).}}
\label{amplitude3d}
\end{figure}


The solution of equation (\ref{eq:R2}) is plotted in figure 1 for $d=3$. We use a normalization which is independent of the temperature $t$: with $\tilde R^2= {R^2 / (-t)}$,  $\tilde T= {T /( -t V_0)}$ and  $\tilde \sigma= {-t \sigma}$. The free energy $F$, calculated as: $F = \int dR \ {\partial F/ \partial R}$, has three branches. The crossing point of the two lower branches determines the location of the first order transition, marked by a vertical dashed line.
The consequence is a first order transition for
\begin{equation}
\tilde \sigma_c = - \sigma t_c  {\ \lower-1.2pt\vbox{\hbox{\rlap{$<$}\lower5pt\vbox{\hbox{$\sim$}}}}\ } 4.5,
\label{equ1erordre}
\end{equation} i.e. the mean amplitude $R^2$ jumps at $T_c$. For $d = 2$, we have also a first order transition for $\tilde \sigma_c {\ \lower-1.2pt\vbox{\hbox{\rlap{$<$}\lower5pt\vbox{\hbox{$\sim$}}}}\ } 1.25$. For $d=1$, there is no transition.

A good approximation for the critical temperature due to phase fluctuations is given by (\ref{equationTc}) with $ \langle R^2 \rangle = -t_c$: 
   \begin{equation}
   t_c \approx -{ T_c / ( \omega V_0)}
  \label{equtc}
  \end{equation}
Combining this last equality and (\ref{equ1erordre}), the first order transition occurs then for  $ \sigma {T_c / V_0} {\ \lower-1.2pt\vbox{\hbox{\rlap{$<$}\lower5pt\vbox{\hbox{$\sim$}}}}\ } C $
where $C = \omega \ 4.5 \approx 9.9$ for $d=3$.\\


What are the influence of amplitude fluctuations around the saddle point that we found? Do they destroy the first order transition or not? To answer to this question, we include harmonic amplitude fluctuations in our computation. We separate phase and amplitude, introduce three variational parameters and derive two self consistent equations for these parameters.

 The effective potential $U_{eff} := \sigma \left( t  R_i ^2  +  {1\over 2} R_i^4 \right) - {T \over V_0} \log R_i $ of the effective hamiltonian (\ref{Heff}) has only one minimum for all temperatures. So one can expect that a gaussian approximation for the amplitude gives a good approximation for all temperatures. We want also to get an hamiltonian with no direct coupling between phase and amplitude, but with effective constants that keep the memory of their interaction. The idea is then to separate phase and amplitude as:
\begin{equation}
R_i ^2  f_i \qquad  \rightarrow \qquad    R_i ^2 \langle f_i \rangle+\langle R_i ^2\rangle  f_i
\label{separation}
\end{equation}
Therefore, we set the {\bf trial hamiltonian}:
\begin{equation}
H_{t}[R,\phi]= \sum_{i}  \left[ B \ (R_i -R_0 \right)^2 +  {1\over 2}  \left( \nabla R_i  \right)^2+  A \  f_{i}  ]
\end{equation}
Using the Bogoliubov inequality, we have $F \leq F_t + \left\langle H_{eff} - H_t \right\rangle_t = \tilde F$, where $F$ is the free energy and $\langle ...\rangle_t$ is the canonical average with respect to $H_t$. The right hand side is thus to be minimized with respect to the constants $A$, $R_0$ and $B$ to give the best approximation of $F$. We introduce also the local amplitude fluctuation $\eta_i= R_i-R_0$. The derivative  of $\tilde F$ with respect to these parameters gives three equations:
\begin{eqnarray}
& A  =  \langle R^2 \rangle      \label{} \\
&  \left[ \sigma (t+3  \langle \eta^2 \rangle ) + \left\langle f \right\rangle  \right]   R_0^2+ \sigma R_0^4- {T \over V_0} {R_0 \over 2} {\partial \langle \log R \rangle \over \partial R_0}  =  0   
\label{equationA} \\
 & \sigma (t+3 \langle \eta^2 \rangle ) + \left\langle f \right\rangle  - B  + 3 \sigma R_0^2- {T \over V_0}  {\partial \langle \log R \rangle \over \partial B }  = 0   
\label{equationB}
\end{eqnarray}
where all indices $t$ and $i$ are dropped. $\langle \log R_i \rangle$ is computed by a cumulant expansion. Equations (\ref{equationA}) and (\ref{equationB}) are the central result of this work. The mean square amplitude fluctuation is
$$
  \langle \eta^2 \rangle =  {T \over V_0}1/V \sum_{|\vec k|< \Lambda}  {1 \over 2} {1 \over B +{k^2 / 2}} 
$$
where we extend the lower bound $-R_0$ to $- \infty$ assuming that amplitude fluctuations are small.  $\Lambda$ is the reduced cut-off parameter and is computed on the first Brillouin zone.\\ Equation (\ref{equationA}) reduces to equation (\ref{eq:R2}) when the additional amplitude fluctuations are zero: $\langle \eta^2 \rangle = 0$.  The phase diagram is approximately the same as the one without amplitude integration, except that the critical temperature is smaller due to the additional harmonic amplitude fluctuations. The transition is first order for $\sigma  {\ \lower-1.2pt\vbox{\hbox{\rlap{$<$}\lower5pt\vbox{\hbox{$\sim$}}}}\ }  9.9 \ V_0/T_c$ for $d=3$.

\begin{figure}
\let\picnaturalsize=N
\def\picsize{8 cm}
\def\picfilename{figGL2.eps}
\ifx\nopictures Y\else{\ifx\epsfloaded Y\else\input epsf \fi
\let\epsfloaded=Y
\centerline{\ifx\picnaturalsize N\epsfxsize \picsize\fi \epsfbox{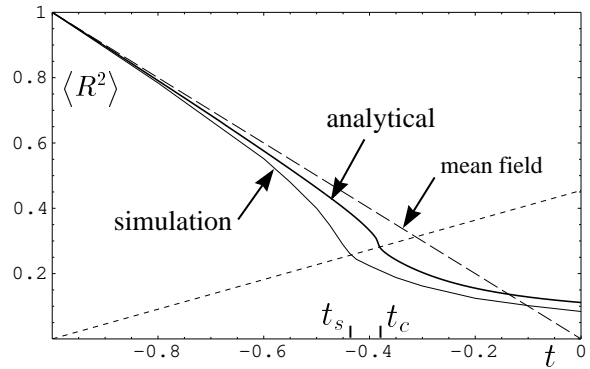}}}\fi

 \caption{{  {\bf Expectation value of $R^2$ for $\bf d = 3$ in the continuous regime.}  The thin line shows the simulation. We set the analytical critical temperature $t_{c} \approx -0.38$ and the one from the simulations $t_s \approx -0.45$.} \label{amplitude3dfluc} }
\end{figure}

In figure \ref{amplitude3dfluc}, we compare the simulations of Nguyen and Sudb\o \ {\cite{nguyen}} with our phase-amplitude separation method for a corresponding $\sigma/(V_0/T_0) = 18$ (continuous regime). We see that the simulated  $\langle R^2 \rangle$ shows an inflexion point at $T_c$ as the analytical $\langle R^2 \rangle$ does. This is perhaps the forerunner for a jump at a smaller $\sigma$. Both methods agree with the fact that there is no first order transition at large $\sigma$. To our knowledge, accurate simulations for smaller $\sigma$ are not avalable.

 A remarkable thing is the fact that a negative $\sigma$ produces a first order transition (\cite{kleinert} p.340). Therefore, the critical line which is believed until now to separate the region $\sigma < 0$ and $\sigma > 0$ is pushed in the region with $\sigma > 0$ by phase fluctuations. The limiting case $\sigma < 0$ is then consistent with a first order transition for $\sigma  {\ \lower-1.2pt\vbox{\hbox{\rlap{$<$}\lower5pt\vbox{\hbox{$\sim$}}}}\ }  9.9 \ V_0/T_c $. We mention here that a variational approximation due to Halperin {\em et al} \cite{halperin} for gauge field fluctuations yields an equation that is very similar to equation (\ref{eq:R2}) except that our function $f(K) = \langle f_i \rangle$ due to phase fluctuations is replaced by the expectation value $\langle \vec{A} ^2 \rangle$. For $d=3$, magnetic fluctuations of the gauge field $\vec{A}$ produce a first oder transition and move the critical point into the positive $\sigma$ region. The domain of validity is also in the small $\sigma$ region as for our approach.
 
We now show that the origin of the first order transition of equations (\ref{equationA}) and (\ref{equationB}) is not due to the approximation itself and that the domain of validity of the approximation is precisely in the first order domain for $d=3$. We establish a quantitative criterion by comparing the size of the critical gaussian region according to the Ginzburg criterion  with the critical temperature due to phase fluctuations. 
The Ginzburg criterion \cite{goldenfeld} measures the importance of correlated gaussian fluctuations that are mainly due to the amplitude, i.e. they are negligible if
$$
|t| \ >> \ t_{G} := \left({T_c \over 2 V_0} \right)^{2 \over 4-d}    \sigma^{d-2 \over 4-d}
$$
where $t_G$ defines the critical region. $|t_c|$ from equation (\ref{equtc} is a measure of the importance of phase fluctuations. Our approach is then valid when (amplitude) fluctuations are dominated by phase fluctuations, i.e. when $|t_c| >> t_G$. Setting $ \omega \approx d - 1$ which is almost the correct XY critical temperature, we get the criterion:
\begin{equation}
{d-1 \over 2} \left(\sigma  {   T_c \over 2 V_0 } \right)^{d-2 \over 4-d} << 1
\label{criterion}
\end{equation}
 For $d=1$, the criterion is in agreement with our results (not shown here) which are almost identical to the exact solution \cite{scalapino}. For $d=2$, the criterion is not clear. The approximation is reliable precisely in the first order domain for the 3$d$ case ($\sigma T_c /V_0 << 1 $), and everywhere for $d=4$. For $d >4$, the domain of validity is in the continuous domain. Therefore, we conclude that the first order transition is not a consequence of the approximation for $d=3$, whereas a doubt remains for $d=2$.

The first order transition is due to the fact that the energy of
vortices can be lowered by a reduced value of the amplitude, which is
energetically favorable when the potential energy of the amplitude is
sufficiently soft. In this case, the usual transition scenario, given by
an unbinding  of topological excitations, seems to be replaced by a
sudden proliferation of the latter. Such a process may be difficult to describe within the renormalisation group $\varepsilon$-expansion where the interplay between amplitude and phase is not explicitely taken into account \cite{noteRG}. A similar scenario has been found by Minnhagen {\em et al.} \cite{minnhagen} for the 2$d$ Coulomb gas. They show that a large vortex fugacity $y= \exp (-\beta \gamma |\psi|^2 \pi^2/2 )$ can lead to a discontinuous transition produced by a proliferation of vortices, whereas small fugacity causes the usual Kosterlitz-Thouless transition. When $\tilde{\sigma}$ is small, our scenario corresponds to the case of large vortex fugacity. The mean amplitude can become much smaller than the one of the XY model (see figure 1 where $ \tilde{R} =1$ corresponds to the XY model). For large  $\tilde{\sigma}$ ($\tilde{R}  \approx 1$) , the KT transition is recovered.
Our result ($\tilde{\sigma}_c \approx 1.25$) is in good agreement with the one of Bormann and Beck \cite{bormann} who found a first order transition for $\tilde{\sigma}  {\ \lower-1.2pt\vbox{\hbox{\rlap{$<$}\lower5pt\vbox{\hbox{$\sim$}}}}\ } 1$.

For real superconductors, $\sigma$ is smaller than 1 \cite{bormann}. Therefore, superconductors with low $T_c / V_0$ could have a first order transition. A candidate for a possible observation would be a 3$d$ superconductor with a critical region $t_{G}$ that is not too small but still in the first oder domain. In the BCS limit, the transition becomes very weakly first order such that there is no contradiction with the continuous behaviour of BCS superconductors. Underdoped cuprates are quasi 2 dimensional and have a low $V_0/T_c$. Their transition is then XY-like, whereas overdoped cuprates, that are almost 3$d$ and have a large  $V_0/T_c$, could have an observable first order transition. It is, however, interesting to remark that measurements of the entropy change $\Delta S$ at the vortex lattice melting transition of Bi$_2$Sr$_2$CaCu$_2$O$_8$ \cite{morozov} have shown a dramatic increase in $\Delta S$ per vortex when the zero field transition is approached. This could be a hint that the superconducting transition remains first order even in zero field.\\

 In this paper we have investigated the thermodynamic properties of the
classical Ginzburg-Landau (GL) model. It is determined by two model
parameters, $\sigma$ and $V_0$.  $\sigma$ governs the strength of amplitude fluctuations
and $V_0$ the overall strength of fluctuations of the complex GL field. We have treated the model by a variational approximation which takes into
account the coupling between phase and amplitude through effective
coupling constants. Minimizing the corresponding variational free energy
leads to a set of self-consistent modified GL equations containing phase
and amplitude fluctuations. The behaviour of the GL transition changes when the ratio
$\sigma/(V_0/T_c)$ is varied: for $\sigma  {\ \lower-1.2pt\vbox{\hbox{\rlap{$<$}\lower5pt\vbox{\hbox{$\sim$}}}}\ } C \ V_0/T_c$ the transition of first order with $C=9.9$ for $d=3$ and  $C=2.2$ for $d=2$.

For $d=3$, we showed that phase fluctuations dominate the transition in the first order domain and amplitude fluctuations can be neglected. We thus conclude that a first order transition is indeed a valid scenario for the GL model, once the amplitude of the field can sufficiently adapt in order to lower the total energy of the system.\\

 We thank D. Baeriswyl, X. Bagnoud, M. Dzierzawa, C. Verdon, H. Fort and C. Wirth for very useful discussions. This work has been supported by the Swiss National Science Foundation (project No. 2000-056803.99).


\begin{references}
{\small
\bibitem{gorkov} Gorkov L P, {\em Zh. Eksperim. i Teor. Fiz.} 36, 1918, (1959)

\bibitem{randeria} Randeria M, {\em Bose Einstein Condensation}, Griffin A, Cambridge UP, (1995)

\bibitem{simanek} Simanek E, {\em Inhomogeneous Superconductors}, Oxford UP, (1994)

\bibitem{pelcovits} Pelcovits R A, Ph. D. Thesis, Harvard University

\bibitem{toner} Toner J J, Ph. D. Thesis, Harvard University

\bibitem{wilson} Wilson K G {\em Rev Mod Phys} 55, 583, (1983) 

\bibitem{wilsonfisher} Wilson K G, Fisher M E {\em Phys Rev Lett} 28, 240, (1972) 

\bibitem{bormann} Bormann D, Beck H, {\em J Stat Phys} 76, 361, (1994) 

\bibitem{minnhagen} Minnhagen P, Wallin M, {\em Phys Rev B} 28/9, 5378,(1983) 

\bibitem{BCS} Bardeen J, Cooper L N, Schrieffer J R, {\em Phys Rev} 108, 1175, (1957)

\bibitem{halperin} Halperin B I, Lubensky T C,  Ma S-k,  {\em Phys Rev Lett} 32, 292 , (1974)

\bibitem{emery} Emery V J, Kivelson S A, {\em Nature} 374, 4347, (1995)

\bibitem{kleinert} Kleinert H, {\em Gauge Fields in Condensed Matter} (1980)

\bibitem{goldenfeld} Goldenfeld N, {\em Lectures on Phase Transitions and Renormalisation Group}, Addison-Wesley, (1992)

\bibitem{nguyen} Nguyen A K, Sudb\o \ A, {\em cond-mat}/9907385, (1999)

\bibitem{scalapino} Scalapino D J, Sears M, Ferrell R A, {\em Phys Rev B}  6/9, 3409, (1972)

\bibitem{noteRG} The first order transition found, for example, in \cite{wilsonfisher} is related to the cubic anisotropy in the fourth order coupling rather than to the interplay between amplitude and phase for the isotropic model treated here.

\bibitem{morozov} Morozov N, Zeldov E, Majer D, Konczykowski M, {\em Phys Rev B} 54, R3784,(1996) 

}

\end{references}
\end{document}